\documentclass[pra,aps,showpacs,preprint]{revtex4}
\usepackage{dcolumn}
\input epsf
\usepackage{epsfig}

\def\km{{\rm km}}

\begin{document}

\title{Neutrino telescopes' sensitivity to dark matter}
\author{Ivone F.\ M.\ Albuquerque}\thanks{Electronic mail: IFAlbuquerque@lbl.gov}
\affiliation{Space Science Laboratory and Astronomy Department, 
        University of California, Berkeley, California \ 94720 }

\author{Jodi Lamoureux}\thanks{Electronic mail: JILamoureux@lbl.gov}
\affiliation{National Energy Research Scientific Computing Center, Lawrence Berkeley 
National Laboratory, Berkeley, CA 94720.}

\author{George F.\ Smoot}\thanks{Electronic mail: GFSmoot@lbl.gov}
\affiliation{Lawrence Berkeley National Laboratory, Space Sciences Laboratory and
Department of Physics, University of California, Berkeley, CA 94720.}

\date{25 Apr 2002}

\begin{abstract}
The nature of the dark matter of the Universe is yet unknown and most likely
is connected with new physics. The search for its composition is under way through
direct and indirect detection. Fundamental physical 
aspects such as energy threshold, geometry and location are taken into account to 
investigate proposed neutrino telescopes of km$^3$ 
volume sensitivities to dark matter. These sensitivities are just sufficient
to test a few WIMP scenarios. Telescopes of km$^3$ volume, such as IceCube, 
can definitely discover or exclude superheavy (M$>10^{10}$ GeV) 
Strong Interacting Massive Particles (Simpzillas). Smaller neutrino telescopes such as
ANTARES, AMANDA-II and NESTOR can probe a large region of Simpzilla parameter space.
\end{abstract}

\pacs{13.15.+g,95.35+d,98.80.Cq}

\maketitle

\section{Introduction}
Most of the Universe's matter is non luminous. There are many
predictions for its composition but the nature of dark matter is yet unknown.
The possibility of accounting for this matter within the standard model of particle 
physics is small and is constrained both by particle physics and by
cosmology. Discovery of the dark matter composition will probably reveal
new physics. 

Many models based in physics beyond the standard model 
propose solutions to the dark matter problem. Most of
these predict the existence of new particles. These new particles can be detected 
directly through nuclei recoil 
or through secondary products of their annihilation. 
The decay of particles produced in
the annihilation or decay of these new particles generates a flux of
high energy neutrinos. The direct and the secondary neutrino signatures are
complementary one to another. Detection -- or lack of detection --
of the secondary neutrinos is therefore important for understanding
the nature of dark matter as well as for uncovering the new physics. 

A leading dark matter candidate is the Weakly Interacting Massive Particle
(WIMP). A thermal stable relic from the early Universe, its abundance is 
inversely proportional
to its thermal averaged annihilation cross section $<\sigma v>$. 
If this cross section 
is estimated from the weak-scale interactions, the resulting 
abundance is close to the observed energy density of dark matter \cite{jung}. 
This coincidence makes WIMPs a strong candidate for the non luminous-matter.

However WIMP is a broad category and its identity and character are yet unknown.
The most investigated candidate is the neutralino, proposed as the lightest 
supersymmetric particle (LSP) within the minimal supersymmetric standard model (MSSM).
If neutralinos exist and compose the dark
matter of the Universe, they should be captured in the Sun and in the
Earth. Once captured they should concentrate near the center and 
annihilate with each other into particles which would produce high energy 
neutrinos. The flux of neutrinos and charged current induced muons arriving in neutrino
telescopes has been estimated in several analysis 
\cite{je,bek,beg97,beg98,jonathan}. 

Another dark matter candidate is a Strongly Interacting Massive Particle
(SIMPs). Capture and annihilation of superheavy (mass above $\sim 10^{10}$ GeV) 
SIMPs (Simpzillas) in the Sun or in the Earth
will also produce secondary high energy neutrinos \cite{ahk}. 

Simpzillas represent one category of superheavy relic dark matter (another category
can, for instance, be composed of superheavy weakly interacting massive particles).
The neutrino flux from superheavy relic dark matter under the assumption that the
decay or annihilation of these particles is the source for the ultra high energy
cosmic rays (UHECR) has been estimated in \cite{francis}. UHECR
are events which violate the so called GZK limit \cite{gzk}. This 
implies that the UHECR flux normalizes the the superheavy relic decay or 
annihilation rate. Under their assumptions for the annihilation 
of these particles AMANDA-II results \cite{amnature} makes it 
unlikely that they produce the observed UHECR without some method of neutrino
suppression \cite{francis}. However different assumptions \cite{sarkar}
predict a lower neutrino flux which still allows the possibility that superheavy relic
dark matter are the source of UHECR.

After summarizing predictions of the muon rate arriving at neutrino telescopes from
WIMPs, we will review the estimate for the flux of charged current induced taus from 
neutrinos produced in the annihilation of Simpzillas in the Sun \cite{ahk}.
We then estimate the detection rate of current or proposed
detectors such as AMANDA-II, ANTARES, IceCube and NESTOR. We also consider a
km-scale detector in the Mediterranean sea 
which would correspond to a larger version of ANTARES or NESTOR. In estimating the
detection rate we
determine the optimum detector energy threshold for indirect detection of Simpzillas.

We show that when the fundamental physical aspects of neutrino telescopes are
considered discovery of dark matter is assured, if it is composed of strongly interacting
superheavy particles. The sensitivity of proposed km$^3$ volume detectors is only
able to probe a few WIMP models in a currently very large phase space of models.

In the next section we summarize the estimates for the muon rate at neutrino
telescopes based on predictions for WIMP annihilation \cite{beg98}. 
Also summarized is the muon rate based on the constrained MSSM (CMSSM). We then discuss 
the possibility of WIMP discovery by neutrino telescopes.

In Section~\ref{sec:simp} we summarize the expected event rate from Simpzilla annihilation
in the Sun \cite{ahk}. In Section~\ref{sec:simprate} we include basic physical aspects of
neutrino telescopes namely energy threshold and km$^3$ detection volume and 
estimate the tau event rate for contained events. In Section~\ref{sec:simpdet}
we show the reach of current detectors. The conclusion
follows.

\section{WIMPS - Neutralino}
\label{sec:wimp}

As discussed above, WIMP is a promising dark matter candidate. Within the WIMP
category, neutralinos -- the LSP -- is the most investigated.
The search for neutralino signature is a search not only for dark matter but also
a test for the supersymmetric extension of the standard model \cite{jung}.

There are mainly two ways to detect WIMPs. One is a direct detection through the
scattering of WIMPs from nuclei. If the halo
of our galaxy consists of WIMPs, they will pass through the Earth. Although their
cross section with ordinary matter is weak and the interactions rare, it is still
possible to measure the small energy liberated from the nuclear recoil due to
a WIMP--nucleus scattering. CDMS \cite{cdms} and GENIUS \cite{genius} are examples 
of experiments that will directly search for WIMP--nucleus interaction.

Indirect detection involves looking for annihilation or interaction products
that propagate through space to the detector.  Here we discuss
detecting high energy neutrinos produced as 
secondary particles from WIMP annihilation. 
Several simulations of this process have been done \cite{je,bek,beg97,beg98} estimating
the rate of charged current induced muons arriving in neutrino telescopes. 
These simulations assume that the neutralino is the WIMP and scans through
MSSM parameters. An analytical
approach has also been developed \cite{jungkam} and allows for a better understanding of the
physical processes such as the capture and annihilation of WIMPs and consequent 
particle production. 

\subsection{High energy neutrino signature from Neutralinos}
\label{sec:wimpbroad}

When WIMPs from the halo pass through
the Sun or Earth material, they may     
lose energy by elastic scattering with nuclei. If their velocities
are reduced below the escape velocity, they are trapped in the Sun or Earth
\cite{gould,press} and further scattering concentrates them together
in the center.  They annihilate with their anti-particles and produce 
leptons, quarks and,
if heavy enough, gauge bosons, higgs and top quarks \cite{jungkam}. Most of these
products will not escape the Sun or the Earth. However neutrinos produced from the
decay of these particles will escape \cite{jungkam}.

Muon neutrinos coming from the Sun or going through the Earth will occasionally interact
with a nucleus and produce muons in a charged current interaction. They will
also have their energy degraded due to a neutral current interaction. The cross sections
for these processes \cite{gandhi} are small but the combination of the density and 
the distance from the center of the Earth or from the Sun is enough to 
allow for significant muon production. After estimating the muon rate  
near the detector one has to account for the effective size
of the detector in order to predict the muon detection rate \cite{als}.

One has also to account for the background. For neutrinos coming
from the Sun the primary background is neutrino production from cosmic ray interactions
in the Solar and Terrestrial atmospheres. 

The indirect neutralino detection rate in neutrino telescopes has been determined
by Bergstrom, Edsjo and Gondolo (BEG) \cite{beg97,beg98}. 
They describe the MSSM parameter range used and consider models
which pass the constraints from accelerator measurements. The relic
neutralino density is determined for each of these models. They
simulate the hadronization and decay of the annihilation products as well as
the interactions of neutrinos in the Earth and on their way out of the Sun.
Neutrino interactions near the detector are also simulated and the muon rate
arriving in the detector is determined. 

Their analysis includes the background from cosmic ray interactions in the Earth and Sun 
atmospheres. In order to reduce this background, 
they use events within an angular
cone of the center of the Earth or of the Sun. They use the method
described in \cite{bek} where it is shown how the angular and/or energy resolution 
of the detector can improve
the sensitivity of neutrino telescopes. However, as in general these telescopes
have poor energy resolution, only the angular resolution 
is used to reduce the background.

\begin{figure}
\centering 
\leavevmode \epsfxsize=230pt \epsfbox{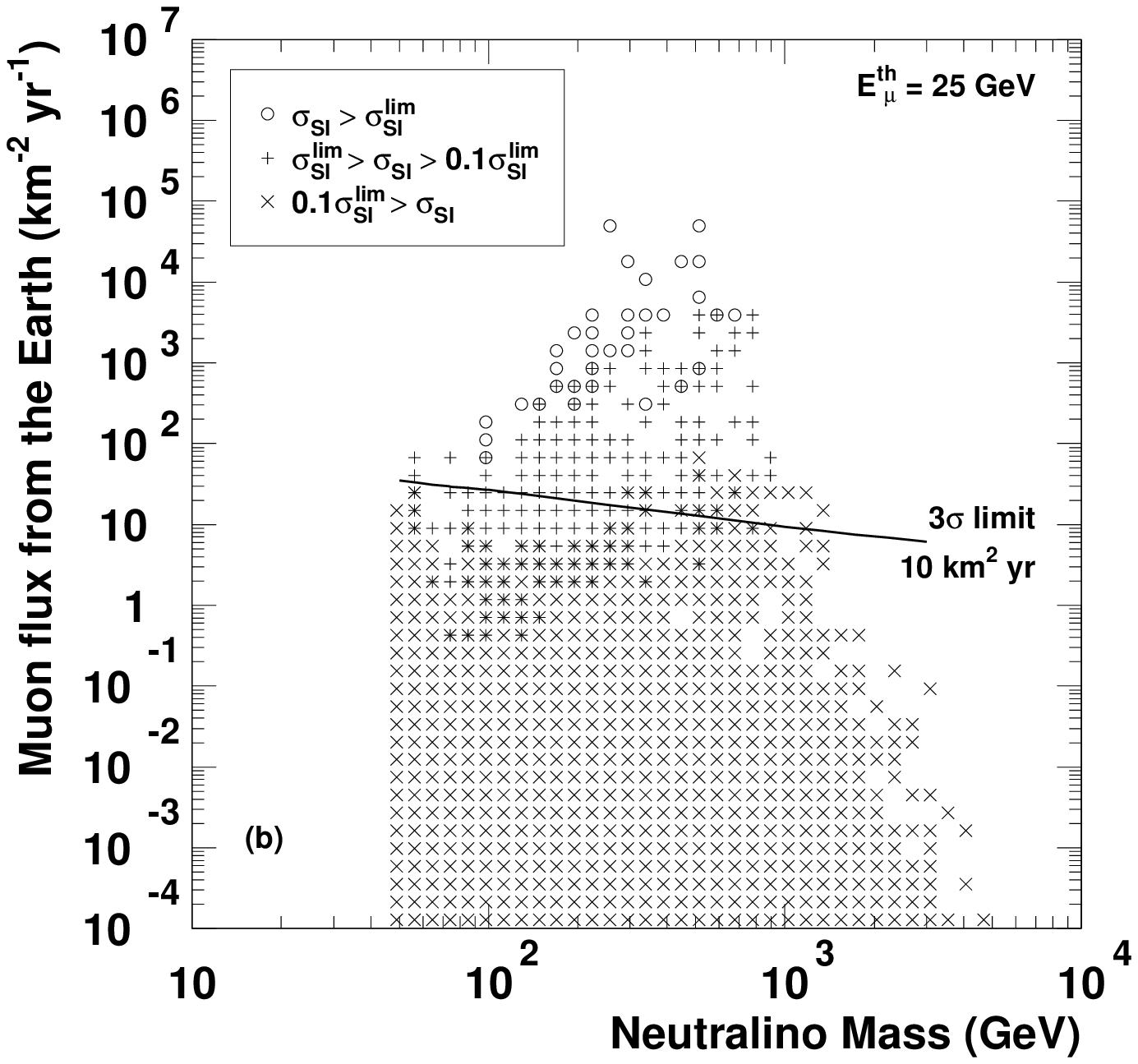}
\epsfxsize=230pt \epsfbox{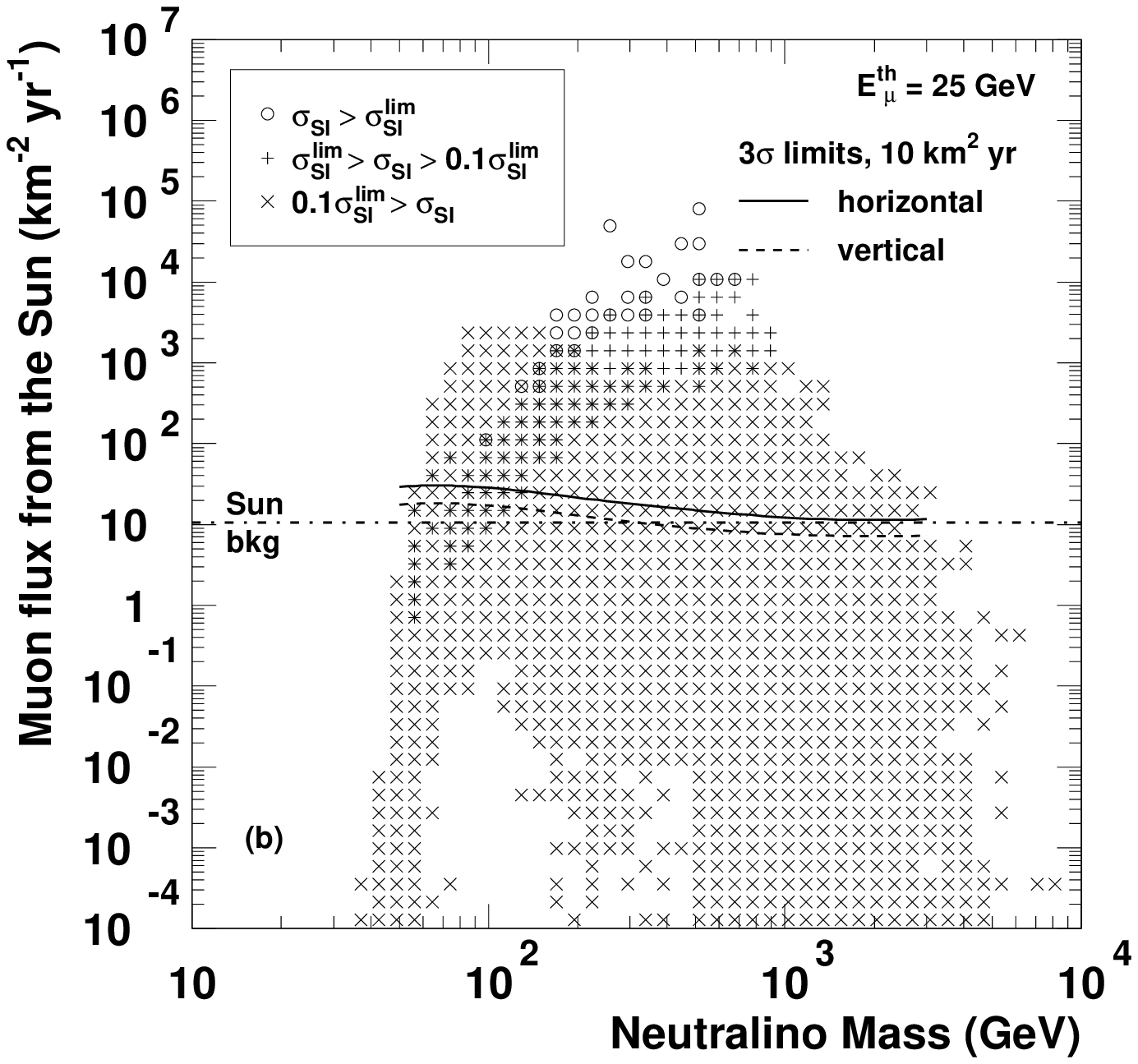}
\caption{The muon fluxes versus the neutralino mass for annihilation in the Earth and
in the Sun obtained by Bergstrom, Edsjo and Gondolo \cite{beg98}. The muon energy 
threshold 
is 25 GeV. The horizontal line 
is a $3 \sigma$ limit to be reached with an exposure of 10 km$^2$ yr. The rate shown 
does not include detector geometry and instrumental
effects. MSSM models excluded
by direct detection experiments are shown with circles and the ones that will be
probed by future direct detection experiments with about a factor of 10 improvement
in sensitivity are shown with a plus sign. (Figure extracted from \cite{beg98}.)}
\label{fig:beg}
\end{figure}

The result from the BEG analysis \cite{beg98} is shown in Figure~\ref{fig:beg}
indicating the estimated
$3 \sigma$ limit for neutralino annihilation in the Earth and in the Sun
obtained for an exposure of 10 km$^2$ year and a 25 GeV energy threshold.
Also shown is the irreducible background from cosmic rays interacting in the Sun.
A similar but reducible background exists for the Earth in atmospherically produced
neutrinos. MSSM models excluded
by direct detection experiments are shown with circles and the ones that will be
probed by future direct detection experiments with about a factor of 10 improvement
in sensitivity are shown with a plus sign. Description of the MSSM models can be found in
\cite{beg98}.

\begin{figure}
\centering 
\leavevmode \epsfxsize=250pt \epsfbox{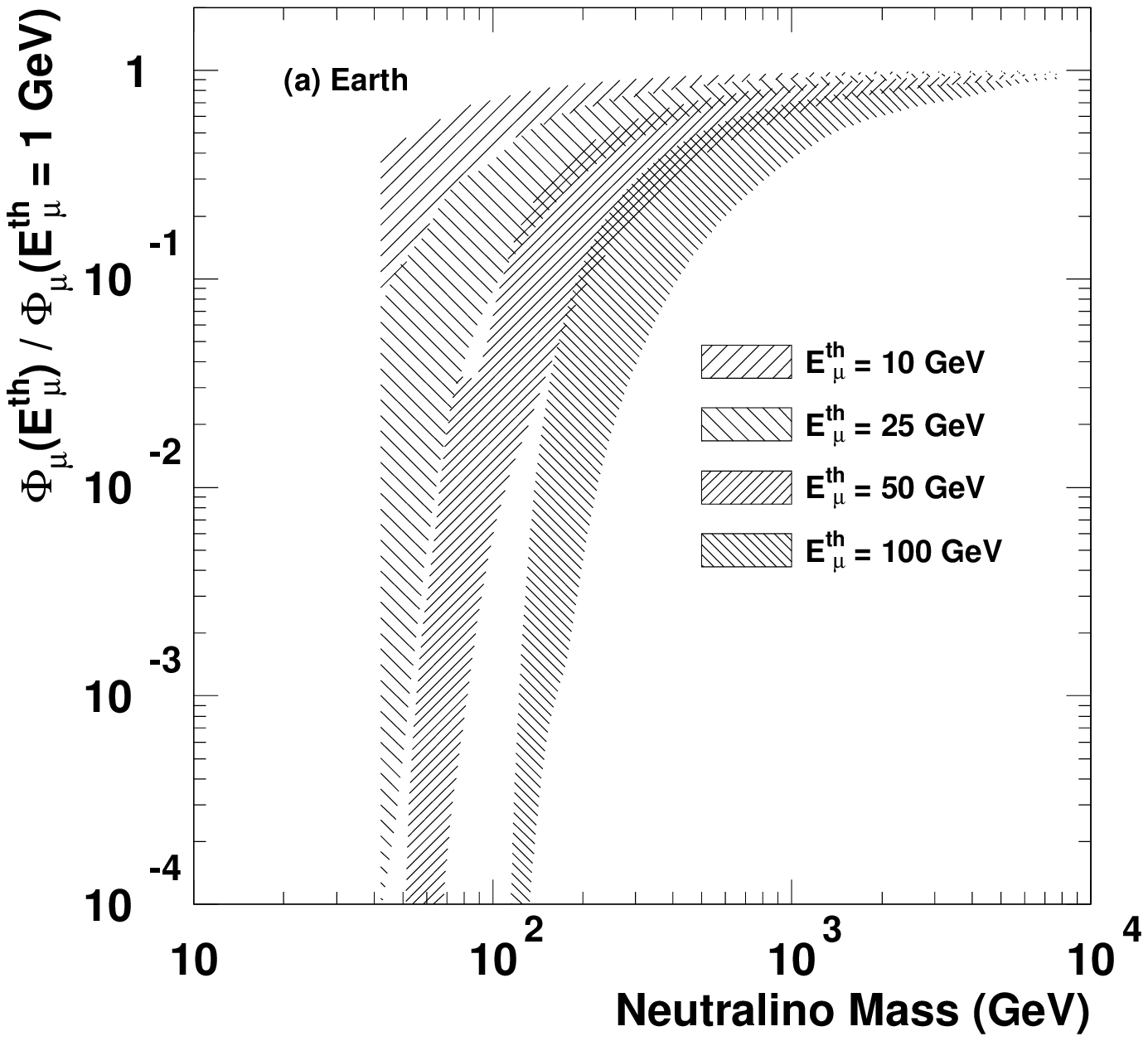}
\epsfxsize=250pt \epsfbox{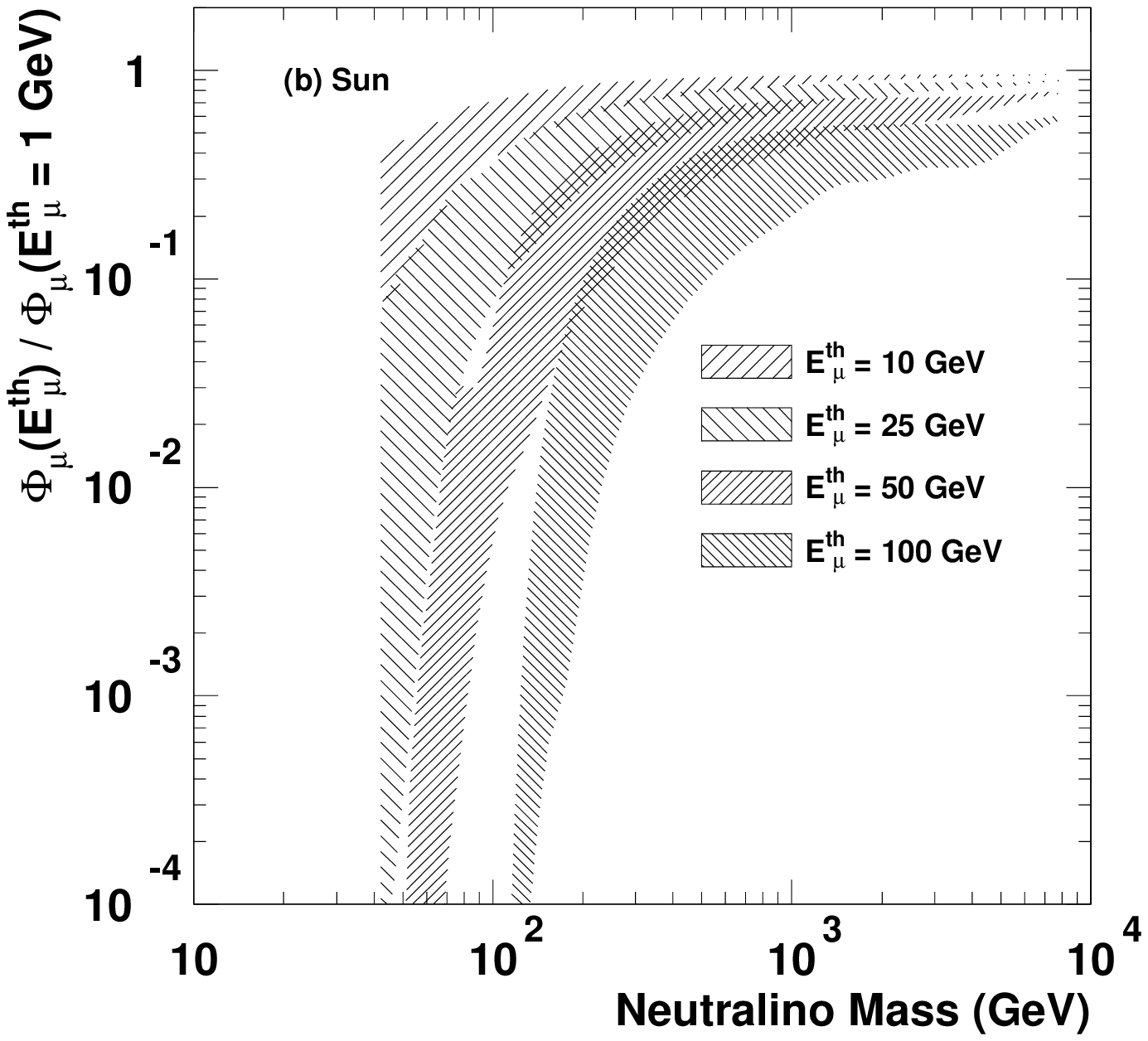}
\caption{The ratio of the muon fluxes with a threshold of \protect${\rm E^{th}_{\mu}}$
to those with a threshold of 1 GeV in (a) the Earth and (b) the Sun versus the
neutralino mass. For each given threshold, a band of allowed ratios is given.
(Figure and caption extracted from \cite{beg98}.)}
\label{fig:begthr}
\end{figure}

BEG \cite{beg98} also show the effect of the detector energy threshold in the muon
detected rates. This is shown in Figure~\ref{fig:begthr}. Different energy thresholds 
are compared to a 1 GeV threshold.
The band represents different degrees of softness of the neutrino spectra for a given 
neutralino mass. It represents the fact that the annihilation branching
fractions are unknown and the resulting neutrino energy spectra might vary
depending on the number of each annihilation product.
A 100 GeV threshold would considerably reduce the signal rate. 

The analysis \cite{beg98} summarized here shows that the potential for 
detection of neutrinos from neutralino annihilation in the Earth is comparable with 
the estimated direct detection sensitivity whereas the
detection from annihilation in the center of the Sun is more promising for neutrino
telescopes. However the muon rate will depend on the detector geometry, energy threshold,
location and also on instrumental effects. In section~\ref{sec:wimpconc} we will expand 
the analysis described above to include these effects. 

\subsection{Neutrinos from constrained MSSM (CMSSM)}
\label{sec:cmssm}

Recently constrains from 
experimental results were used to select MSSM scenarios. These results include searches 
for sparticles,
Higgs boson and $b \rightarrow s \gamma$ decay rate. The CMSSM parameter space also constrains
the supersymmetric relic density to be within the $0.1< \Omega_\chi h^2 < 0.3$ range
which is set by cosmological observations.

A set of parameters within the CMSSM has been chosen 
in order to probe this model in a more systematic way \cite{susybench}. This set was
used to select benchmark models that are representative of a particular set of parameters.
Thirteen models were selected with this purpose.

\begin{figure}
\centering 
\leavevmode \epsfxsize=250pt \epsfbox{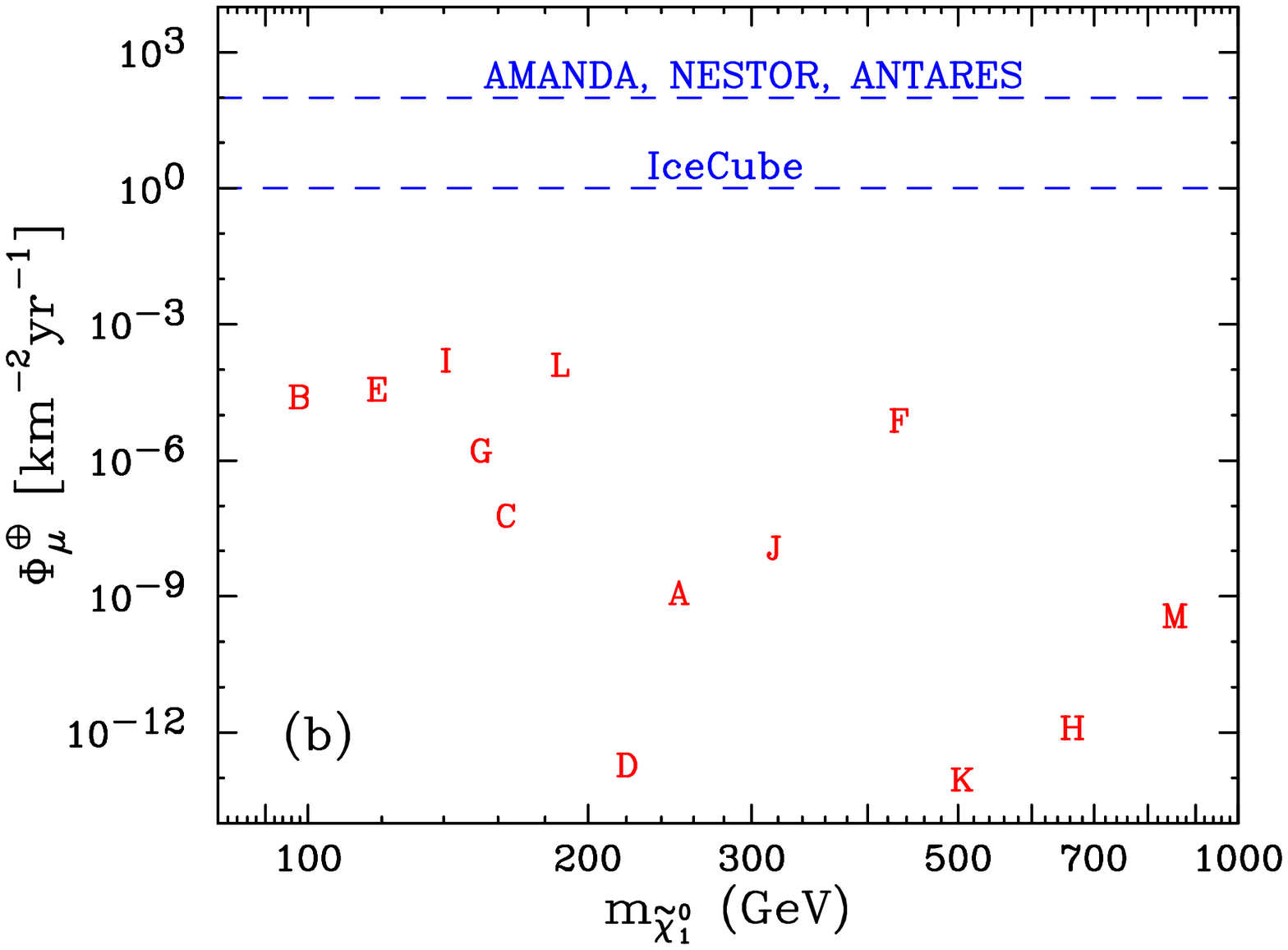}
\epsfxsize=250pt \epsfbox{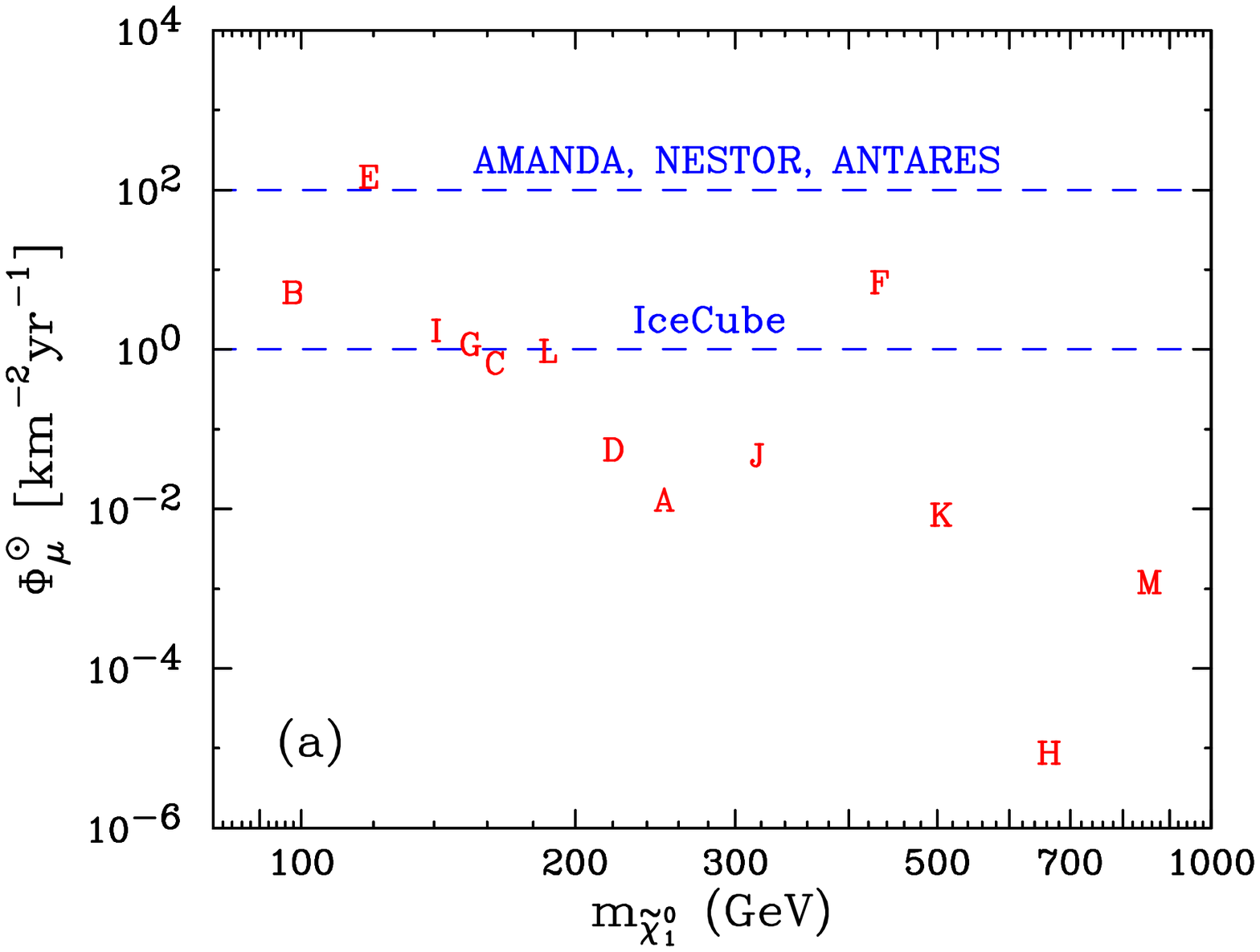}
\caption{CMSSM predictions for neutrino induced muon flux from neutralino annihilation
in the Sun and in the Earth. Each character represents a model within the CMSSM.
Parameters of each set of model are given in \cite{jonathan}. Sensitivities for AMANDA II
\cite{amanda2}, NESTOR \cite{nestor}, ANTARES \cite{antares} and IceCube \cite{ice3} are also
shown. Figure extracted from \cite{jonathan}.}
\label{fig:bench}
\end{figure}

Direct and indirect signatures of the benchmark models are estimated in \cite{jonathan}.
Indirect detection signature is determined assuming neutralino annihilation in the Sun and Earth.
The neutrino charged current induced muon flux from benchmark models is determined and 
their results 
are compared with the quoted sensitivity of neutrino telescopes. These are shown in
Figure~\ref{fig:bench}.

Figure~\ref{fig:bench} also shows the benchmark models that can be probed by indirect detection
assuming neutralino annihilation in the Earth and in the Sun. See \cite{jonathan} for 
description and parameters of each model.
A few representative models can be indirectly tested from neutralino
annihilation in the Sun but not in the Earth. Two of these models (E and F) have exceptional
flux. Both models predict neutralinos with significant Higgsino content which leads to
annihilations into gauge bosons. Gauge bosons will produce hard neutrinos \cite{jungkam}
and the muon rate will be enhanced.

IceCube sensitivity is enough to probe models B, E and F shown in 
Figure~\ref{fig:bench}. 
Models I, G, C and L are on the borderline. 
However some models are below the detector energy threshold.
We will consider this aspect in section~\ref{sec:wimpconc}.

\begin{figure}
\centering 
\leavevmode \epsfxsize=500pt \epsfbox{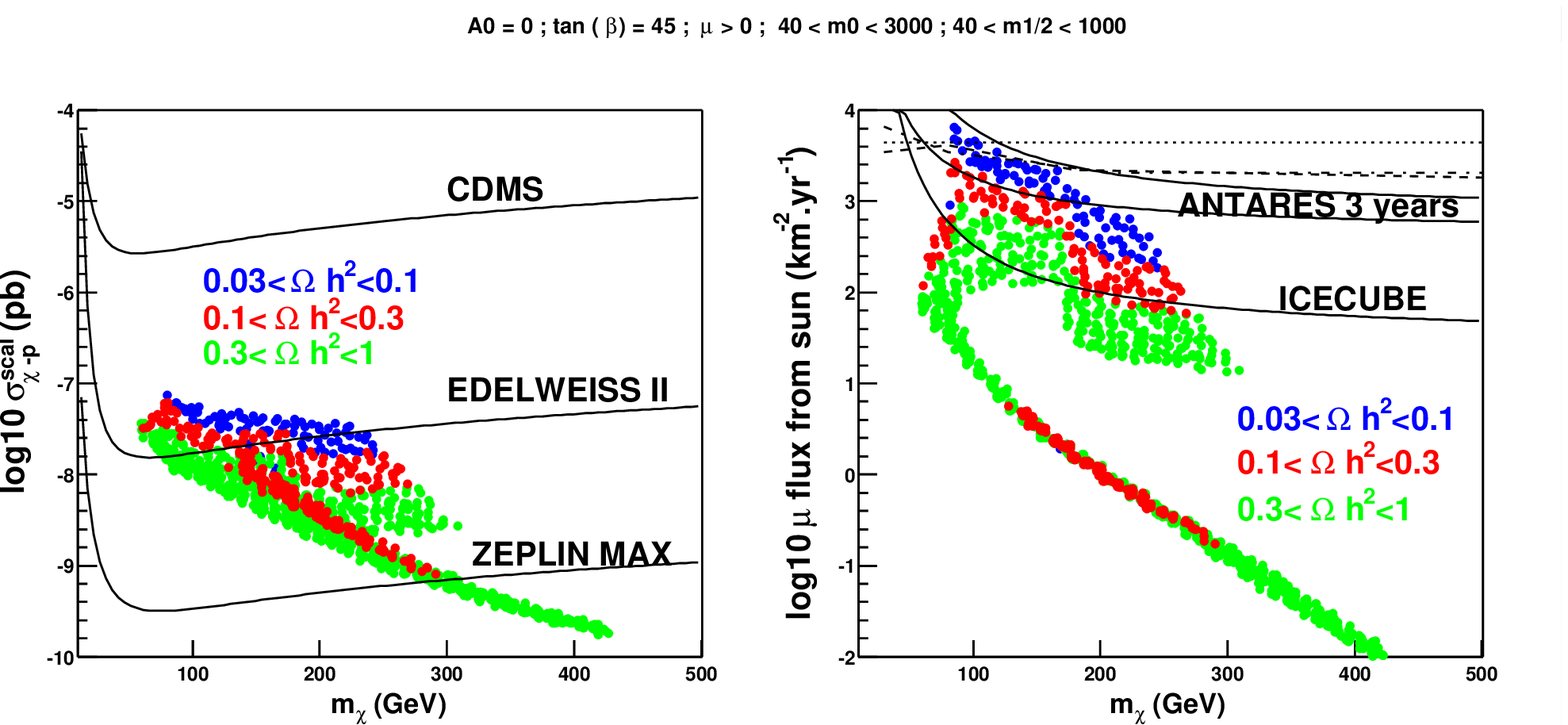}
\caption{Direct and indirect detection experiments sensitivities to neutralinos
assuming a set of mSugra constrained models as determined in \cite{french}. Indirect
detection assumes a 5 GeV detector energy threshold and the rate is shown for annihilation
in the Sun. The dotted, dash-dotted and dashed 
curves in the indirect detection plot are respectively the Macro, Baksan and 
Super-Kamiokande upper limits. The predicted sensitivities for different experiments are
labeled. See \cite{french} for the parameters used. Figure extracted from \cite{french}.}
\label{fig:fr}
\end{figure}

Gravity-mediated supersymmetry breaking in supergravity is also used \cite{sgrav} 
to constrain the MSSM (mSugra model). It allows for a wider range of the cosmological
relic density $0.03 < \Omega h^2 < 0.3$ and is defined by a 5 parameters constrained
MSSM. The muon flux from neutralino annihilation in the Sun from a set of mSugra models
is estimated as well as the direct detection rate.
Their results are shown in figure~\ref{fig:fr}. One problem with this analysis is the low
detector energy threshold which is assumed to be 5 GeV. This will be discussed in
section~\ref{sec:wimpconc}. The conclusion of this analysis is that indirect detection 
of neutralinos assuming the mSugra constraints is beyond reach of
present and future neutrino telescopes for annihilation in the center of the Earth.
For annihilation in the Sun, the region where the neutralino has a larger higgsino
content (which increases the elastic scattering and neutralino annihilation cross sections)
can be probed.

\subsection{Discussion}
\label{sec:wimpconc}

Three scenarios for indirect detection of WIMPs were summarized in this section:
a search for neutrino signature from a broad set of parameters within 
the MSSM, a search for CMSSM benchmarks and a search for models within mSugra. 
In all scenarios, models for WIMP annihilation in the Sun predicts
a higher neutrino rate than annihilation in the Earth.  
We now discuss how the basic physical aspects
of neutrino detectors such as geometry, location and threshold will affect 
these results. 

An analysis of the fundamental physical aspects of instrument performance for
neutrino telescopes is described in \cite{als}. The basic physical characteristics
of an ideal detector are taken into consideration. An ideal detector performance
is translated to include characteristics of
current and proposed detectors. The geometry and location of such detectors
are taken into consideration.

The prediction \cite{beg98} of MSSM scenarios to be tested by neutrino 
telescopes is shown in Figures~\ref{fig:beg}~and~\ref{fig:begthr}. 
These results depend strongly on a low energy threshold.
Also the rates are determined for 10 years of 100\% efficiency in a km$^2$ 
incident area.
If one includes the duty cycle of the detector in a year this will translate to
a longer period in real time. 

The best threshold to be achieved by a km$^2$ incident area detector is set
by the minimum ionizing muon energy loss of about 2.6 MeV/cm.  A 25 GeV 
threshold implies that a 125 meter track was observed.   To determine the track
length, a minimum of a few photomultiplier tubes (PMTs) have to be hit.  Efficient coverage of
a km$^3$ with 5 hits per 125 meters would then require a 64,000 channel detector.
The largest proposed neutrino detector is IceCube with 5,000 channels 
\cite{pdd}. Spreading the
channels over a larger area has the effect of raising the energy threshold
well above 25 GeV.  IceCube will consist of strings with 125 meter string spacing
and 16 meter PMTs spacing on each string.  Simulations of the 
energy threshold
indicate E$^{thr}_\mu > $ 10 GeV for vertical tracks and 200 GeV for horizontal
tracks with something close to 0.7 km$^2$ volume \cite{pdd}.  
Since these predictions do not include background rejection of atmospheric 
muons, it is insightful to note that the AMANDA-II energy threshold for 
muons is 50 GeV \cite{amandaNu, amandaWIMP} which translates into a 
neutralino threshold of 
M$_\chi \mathrel{\vcenter{\hbox{$>$}\nointerlineskip\hbox{$\sim$}}} 100$ GeV.  
The vertical spacing of PMTs in IceCube is similar 
to AMANDA leading us to conclude that a similar threshold may apply. However
as AMANDA and IceCube are located in the South Pole, the neutrinos coming from
the Sun will be arriving at the detector almost horizontally. This implies
that the energy threshold will be closer to 200 GeV which translates into
a higher neutralino threshold of 
M$_\chi \mathrel{\vcenter{\hbox{$>$}\nointerlineskip\hbox{$\sim$}}} 400$ GeV.
In this sense ANTARES is better located since neutrinos from the Sun 
enter the detector at all zenith angles.
A km$^3$ version of ANTARES would be able to perform an analysis 
as a function of the zenith angle.

High energy thresholds reduce
the MSSM parameter space to be tested by neutrino telescopes.
Figure~\ref{fig:begthr} shows that the muon rate
in the detector will be suppressed by about a factor of 10 around 100 GeV and
by a factor of 5 around 1000 GeV when going from a 25 GeV to a 100 GeV threshold.  
In IceCube, limits from the Sun will be several orders of magnitude worse than
shown in Figure~\ref{fig:beg} below a neutralino mass 
of 150 GeV and a few times worse elsewhere due to the detector energy
threshold. 

This together with the fact that for a 25 GeV threshold the background from 
the Sun for neutralino masses around 100 GeV requires  
tens of km$^{2}$ instead of a 10 km$^2$ exposure due to the background from the
solar corona \cite{beg98} lowers the possibility of indirect discovery of WIMPs
in this mass region. At higher energies ($\sim 1$ TeV) the rate
will be suppressed by a factor of 5 due to the higher threshold and reduced
flux. 

The effects of the geometry and location for current and proposed detectors has
been discussed in \cite{als}. The geometrical efficiency of big detectors like
IceCube is close to unity taking into account muons that travel 300 meters or more.  
The primary low energy muon background comes from atmospheric muons 
and muon-neutrinos produced in cosmic ray interactions in the atmosphere.  
Specifically, only up-going
measurements are possible \cite{bek}. WIMPS from the Sun therefore,
may only be detected at night when the Sun is below the horizon.  In the 
South Pole, there is night for 6 months when the Sun dips a maximum of 23.5
degrees below the horizon.  At Mediterranean latitudes, the Sun is below 
the horizon each night.  For both locations exposure times of 10 years 
will actually take 20 years of nightly observing. 

\begin{figure}
\centering\leavevmode \epsfxsize=350pt \epsfbox{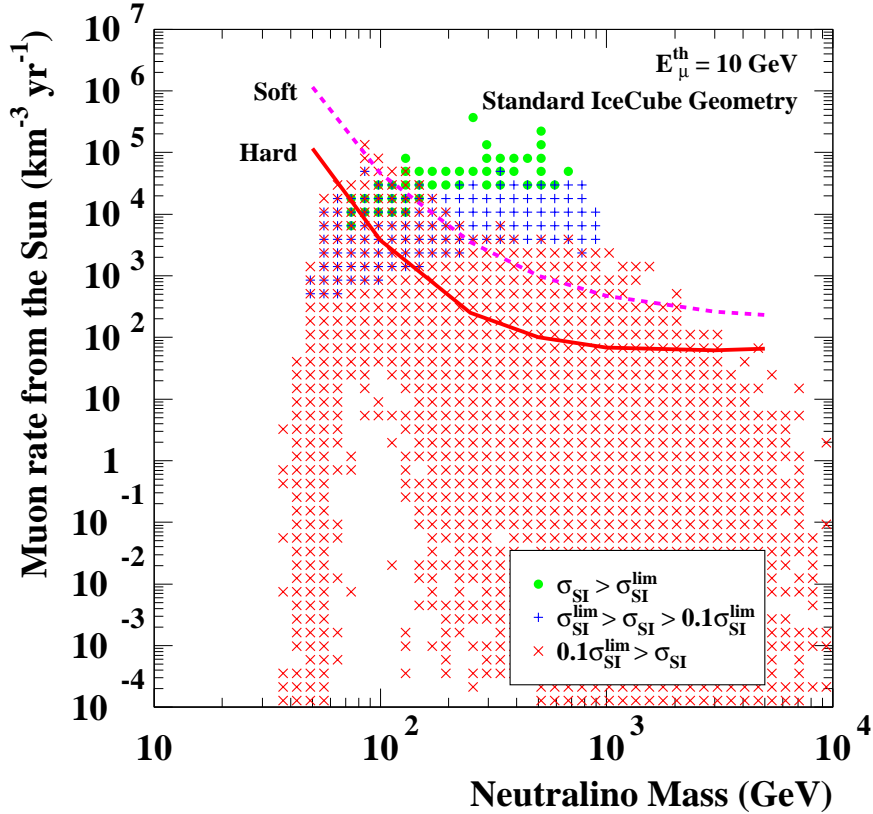}
\caption{Predicted muon rates from WIMP annihilation in the Sun as a function of the
neutralino mass compared to IceCube predicted limit \cite{spier} The IceCube limit is
for 5 years of operation and normalized to a 10 GeV threshold. The ``soft'' represents
WIMP annihilation branching fractions which produce more neutrinos than the ``hard'' 
modes. Figure extracted from \cite{spier}.}
\label{fig:ice3}
\end{figure}

Figure~\ref{fig:ice3} \cite{spier} shows the predicted muon rate from WIMP
annihilation in the Sun and the IceCube predicted limit in 5 years 
live-time (10 years of operation). The values 
shown are normalized to a 10 GeV threshold. The region to be probed
is much smaller than the one shown in Figure~\ref{fig:beg}.

It is also important to note that the direct and indirect detection will be
complementary one to the other. Although a large fraction of the region to be probed 
by indirect detection can also be probed by direct detection, an agreement among these
two techniques is important.

From the 13 CMSSM benchmarks (see section~\ref{sec:cmssm}) \cite{susybench}
one is within the quoted IceCube sensitivity and above its energy threshold (model
F). Four others (I,G,C and L) are on top of the quoted sensitivity which
makes their detection borderline. Models B and E fall below the detector energy threshold.
The others cannot be probed by IceCube.  
An expanded version of ANTARES may
be able to probe models E and B. As ANTARES is located in the Mediterranean,
neutrinos coming from the Sun enter the detector in all angles. For this reason
the energy threshold for neutrinos coming from the Sun is lower than a detector
in the South Pole with same PMT spacing.

For models within the mSugra framework (see section~\ref{sec:cmssm}) \cite{french}
the indirect detection is only possible for neutralino annihilation in the Sun for 
neutralinos with a large higgsino content. The muon flux predicted for a set of these models are
shown in Figure~\ref{fig:fr}. However it was determined for a detector energy threshold of
5 GeV which is too low for a telescope such as IceCube. The results shown for ANTARES
might be better represented since as mentioned above ANTARES energy threshold is 
lower than IceCube, 
though 5 GeV is still not representative of the average energy threshold. The set of
models to be probed by ANTARES is shown in Figure~\ref{fig:fr}.

We conclude this section by stating that the WIMP parameter space is 
very large, and detectors located in the South Pole (as the proposed 
IceCube experiment) or in the Mediterranean (as would be an expanded version of
ANTARES or NESTOR) will probe only those few models with the largest neutrino 
fluxes.
Models with lower fluxes can be probed if the energy thresholds are lowered
and experimental live-time is extended beyond 5 years (10 years of nightly 
observing).
However -- as will be shown in the section that follows -- if dark
matter is composed by superheavy strongly interacting particles their
discovery will be guaranteed for the most natural scenarios. 

\section{Simpzillas}
\label{sec:simp}

It has been proposed by Chung, Kolb and Riotto \cite{rocky} that the dark matter of the
Universe might be composed of a supermassive (mass greater than $10^{10}$ GeV) stable
particle that has never been in thermal equilibrium with the primordial plasma.
In order not to be in thermal equilibrium
the particle annihilation rate must be smaller than the expansion rate. Essentially
the large mass prevents the particle from thermalizing. As it is not a thermal
relic their abundance is not determined by their self-annihilation cross section
but by their mass. These particles can be produced in many ways \cite{rocky}. Among
their production mechanisms are the decay of the inflaton field, gravitational at the
end of inflation or through a broad resonance mechanism of preheating.
The inflaton mass ($\sim 10^{12}$ GeV) is an appropriate scale for their mass
\cite{rocky}.

As the abundance of these superheavy particles depends on their mass and
not in their interaction strength they might interact strongly or weakly with 
normal matter. If they are strongly interacting and
constitute the dark matter of the Universe, they are captured in
the Sun as well as in the Earth as shown in Albuquerque, Hui and Kolb (AHK) \cite{ahk}.
They lose energy while scattering with the Sun or Earth's matter and are trapped in there
once their velocity is smaller than the escape velocity.
The same happens with WIMPS but with one important difference:
WIMPs interact weakly with normal matter and the optical depth of the Sun for these particles
is much less than unity; whereas for Simpzillas, which interact strongly, the
optical depth is much greater than one. The capture and annihilation rate of Simpzillas
in the Sun and in the Earth is estimated in \cite{ahk}. 
The Sun is more efficient in Simpzilla capture and therefore the annilhilation
of these particles will produce a greater neutrino
flux than the one from Simpzilla annihilation in the Earth. For this reason,
we consider the high energy neutrino signature from annihilations in the Sun.

As Simpzillas are ultra heavy particles with masses above $\sim 10^{10}$ GeV, their
annihilation produces hadronic jets which include top quarks \cite{ahk}. The neutrino
rate from top decay in the core of the Sun is estimated as well as the neutrino
emission rate from the Sun.
As shown in AHK \cite{ahk} the rate of muon and electron neutrinos are suppressed since
once they convert into a lepton through a charged current interaction (CC) the lepton will
lose most of its energy before decaying. The same is not true for tau neutrinos since
the tau produced in the CC will immediately decay. As a consequence the tau neutrino
emergent rate is about twice the muon and electron neutrino rates. 
However if neutrino oscillations is taken into account, tau neutrinos might
oscillate into muon neutrinos and vice versa. This effect will be discussed in
section~\ref{sec:osc}. Another enhancement of muon and electron neutrinos are
due to taus decaying close to the edge of the Sun. The tau decays 18\% of the
time into $\nu_\tau \mu \nu_\mu$ and another 18\% into $\nu_\tau e \nu_e$. We call
these $\nu_e$ and $\nu_\tau$ as secondary neutrinos and will take them into
account when discussing oscillations.

\begin{figure}
\centering\leavevmode \epsfxsize=450pt \epsfbox{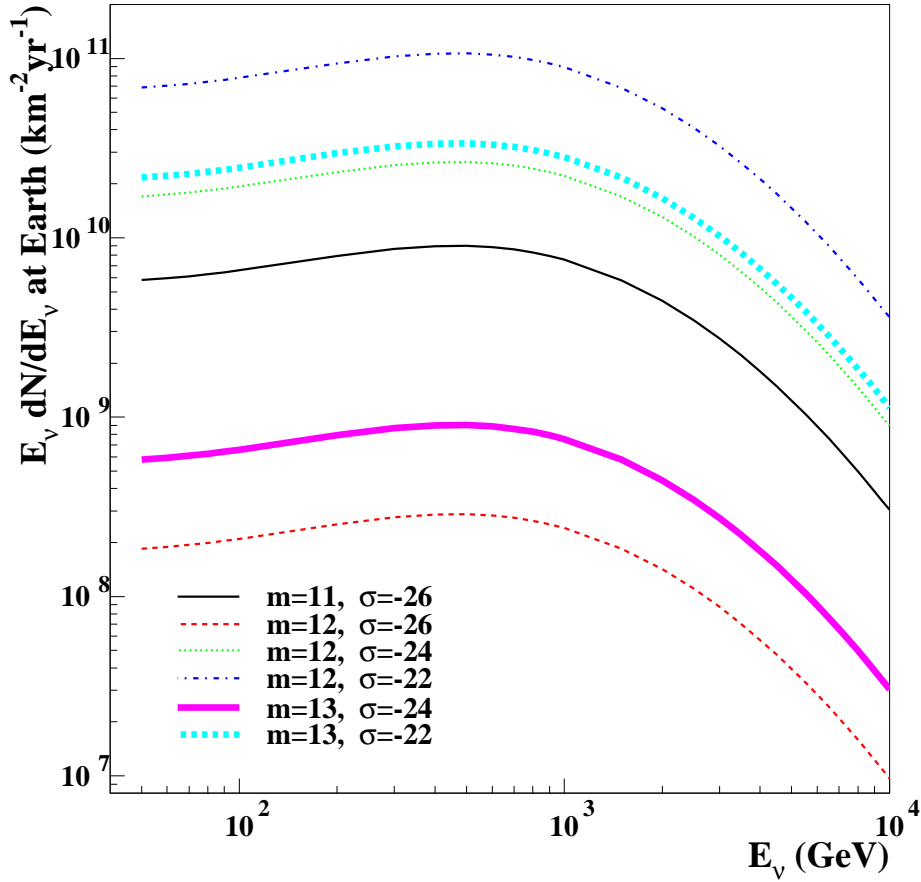}
\caption{Tau neutrino differential rate at the Earth weighted by neutrino energy 
from Simpzilla annihilation in the Sun. Values assumed for
$M_\chi$ and $\sigma_{\chi-N}$ are labeled as $10^{\rm m}$ GeV and $10^{\sigma}$
cm$^2$ respectively. Anti-taus are not included.}
\label{fig:simpterra}
\end{figure}

In section~\ref{sec:simprate} we estimate the sensitivity of neutrino telescopes
to tau neutrinos from Simpzilla annihilation in the Sun. We take the arriving flux
of tau neutrinos at the Earth from the Sun as determined in equation~5.1 of \cite{ahk}.
This rate depends on the Simpzilla mass ($M_\chi$) and the Simpzilla-Nucleon cross section
($\sigma_{\chi-N}$). Figure~\ref{fig:simpterra} shows the tau neutrino differential 
flux arriving at the Earth weighted 
by neutrino energy for different values for $M_\chi$ and $\sigma_{\chi-N}$. This
flux will be enhanced by the anti-tau contribution.
The allowed $M_\chi$ versus $\sigma_{\chi-N}$ space is shown in Figure~\ref{fig:simpallow}.

It is important to note that the results from \cite{ahk} include only neutrinos
produced from top decay. The hadronic jets produced in
Simpzilla annihilation also produce B-mesons which also decay into neutrinos of all 
flavors. They will
therefore enhance the neutrino rate. The shape of the neutrino energy
spectrum from B decays should be the same as that for top decay. Also
the flux shown in Figure~\ref{fig:simpterra} does not include the
anti-neutrinos.

The background for these events consists mainly of atmospheric neutrinos, solar
neutrinos and atmospheric muons. The atmospheric muon background is huge for downgoing events
but 
insignificant for the upgoing ones. As we will show in section~\ref{sec:simprate}
the upgoing event rate from Simpzillas produced neutrinos is sufficiently large
that one does not need to consider the downgoing ones. 

In section~\ref{sec:simprate} we determine the $\nu_\tau$ contained event rate in 
a perfect detector of km$^3$ volume as well as the sensitivity of such detector
for neutrinos from Simpzilla annihilation in the Sun. We include background
estimates. We then estimate the
rate and sensitivity for current or proposed telescopes such as AMANDA-II, ANTARES,
NESTOR and IceCube as well as for a km$^3$ volume detector in the Mediterranean
which would be equivalent to an expanded version of ANTARES and NESTOR.

\section{Tau detected rate from superheavy dark matter}
\label{sec:simprate}
The neutrino flux spectrum from Simpzilla annihilation in the Sun arriving
at the Earth is given by \cite{ahk}:
\begin{equation}
\frac{dF}{dE} = \frac{1}{4\pi D^2}\left(\frac{df}{dE}\right)_{\rm emergent}
=3.5\times10^{-18}\left(\frac{df}{dE}\right)_{\rm emergent} \ \km^{-2} .
\end{equation}
where D is the Sun-Earth distance and the differential in the right is
the neutrino flux emerging from the Sun given in \cite{ahk}.
This flux depends on the Simpzilla--Nucleon cross section and on the Simpzilla
mass. Figure~\ref{fig:simpterra} shows the tau neutrino differential rate
weighted by neutrino energy deduced from this equation. 

This rate has to be weighted by the probability of conversion inside the detector.
This probability is approximately $n \sigma_{CC}(E) L$ where $n$
is the target number density (ice or water for neutrino telescopes), 
$\sigma_{CC}(E)$ is the neutrino-nucleon charged current (CC)
deep inelastic scaterring 
cross section and L is the length of the detector assumed to be 1 km.
We determine $\sigma_{CC}(E)$ using CTEQ4-DIS parton distribution
functions as described in \cite{gandhi}. Although the attenuation of the flux when
going through the Earth is not significant we take it into account
for completeness. This attenuation reduces the initial flux by 
exp($-\int n_\oplus (\sigma_{CC}+\sigma_{NC})dx$) 
where $x$ is the distance the neutrino travels
through Earth, $n_\oplus$ is the Earth number density and $\sigma_{NC}$ is
the neutrino-nucleon neutral current (NC) cross section. 
The event rate spectrum is given by
\begin{equation}
\frac{dR}{dE} = \frac{dF}{dE}\left[n\ \sigma_{CC}(E)\ L \right] A
\ \exp(-\int n_\oplus \left(\sigma_{CC} + \sigma_{NC}\right) dx)
\end{equation}
where $A$ is the area of the detector, assumed to be 1 km$^2$.
It is important to note that the detected rate only includes the contained
events.  We approximate the contained rate by including 
events where the neutrino converts into a lepton
inside the geometrical detector volume.

A $\nu_\tau$ CC interaction with a nucleus will produce a $\tau$ lepton 
with energy $(1 - y )E_\nu$ and a hadronic shower with energy $y E_\nu$, where
$y$ is the fraction of energy transferred to the hadronic shower given
in \cite{gandhi96}. 
Once the tau is produced inside the detector it will almost immediately decay and
produce another $\nu_\tau$ and another shower
(hadronic or electromagnetic). The distance it travels inside the detector is
an important characteristic that can be used at high energy for lepton identification.
This will be discussed in the next section when we describe the double-bang technique
\cite{pakvasa}.

As the tau production implies a simultaneous hadronic shower production and its
decay might also generate another shower inside the detector its signature
is different from one produced by a muon neutrino. The background for taus
will include all interactions that produce a shower inside the detector.
Also, as taus decay almost immediately, they have to be generated inside the
detector. As muons travel a long way before decaying they can be produced outside
the detector and yet be detected. Another important difference is that the
detector angular resolution for a single particle as a muon is much better than for
showers. Neutrino telescopes have angular resolution of about a degree
\cite{barwick} for a long single-particle track and $30^o$ for a shower 
\cite{pdd}. The backgrounds for tau events will be discussed
in the next section as well as ways to reduce them.

\begin{figure}
\centering\leavevmode \epsfxsize=450pt \epsfbox{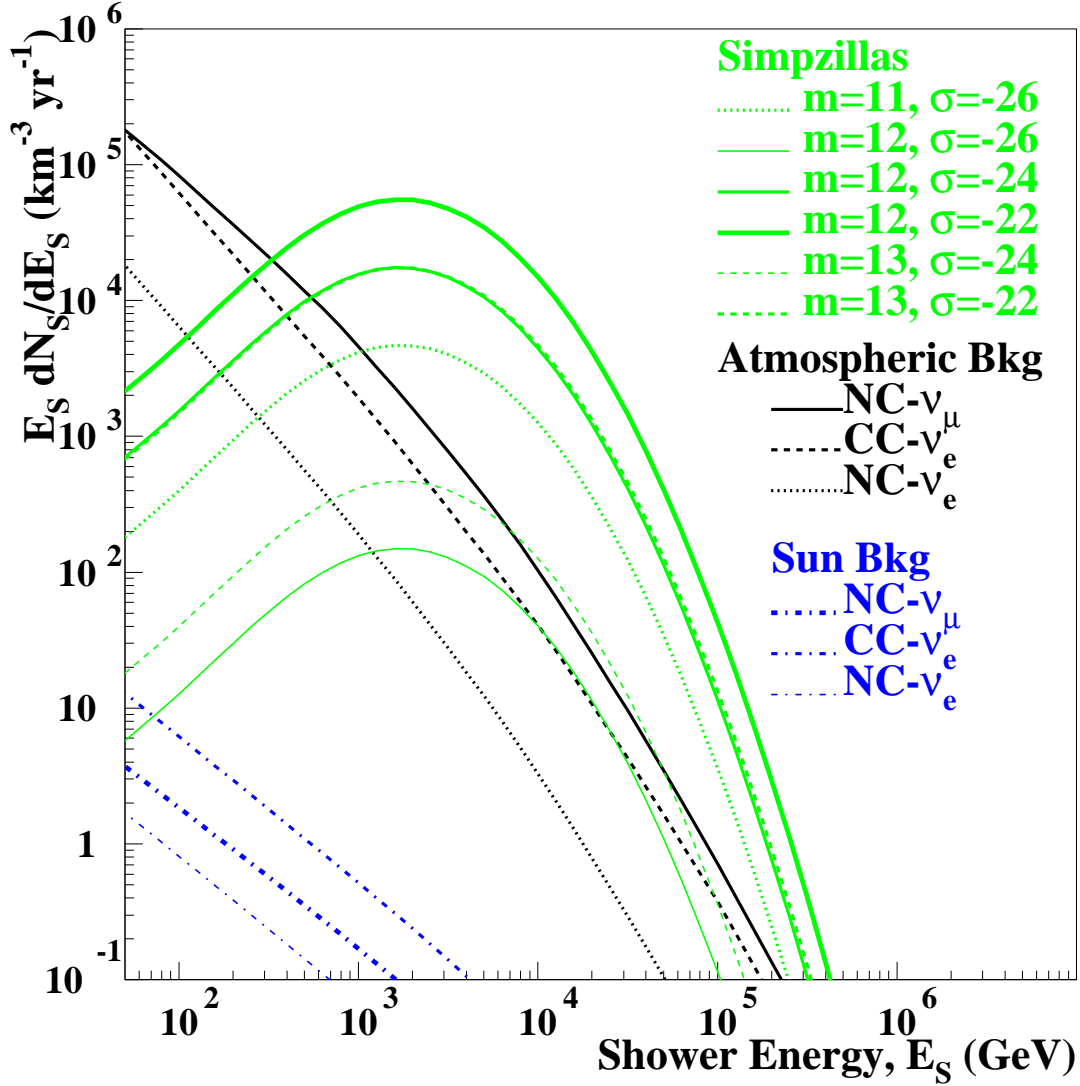}
\caption{Differential shower rate weighted by shower energy. Showers are produced
from tau neutrino charged current interactions. The tau neutrinos are secondary products
of Simpzilla annihilation in the Sun. Different values for $M_\chi$ and $\sigma_{\chi-N}$
are assumed and labeled as $10^{\rm m}$ GeV and $10^{\sigma}$. 
Also shown is the background of showers generated by muon neutrino--nucleon 
neutral current interactions, electron neutrino--nucleon charged and neutral current
interactions from the Earth \cite{volkova} and from the Sun \cite{ingel}. 
The background is shown without any technique to be reduced. Even in this
worse possible scenario the signal can be detected.}
\label{fig:simpcasc}
\end{figure}

In Figure~\ref{fig:simpcasc} we show the estimated event rate from Simpzilla 
annihilation in
the Sun. The rate is for events contained in a km$^3$ volume and includes both taus and 
anti-taus \footnote{from now on when taus are mentioned they include the anti-taus.}. 
The number of showers
generated by a tau neutrino conversion into a tau versus the shower energy is shown.
Since the tau will most likely decay very close to the vertex where the shower
was produced, we consider that 100\% of the tau neutrino energy went into
the shower. 
Also shown are the largest backgrounds, i.e., assuming there is no angular resolution
and taking the background from full sky. The background sources will be discussed in
the next section but as can be seen in this Figure -- even in the worst case -- the event
rate is large and readily detected.

\subsection{Background}

The backgrounds for tau neutrino signature from Simpzilla annihilation in the
Sun are produced by any interaction that generates a shower inside the
detector.  Showers associated with muon events are not considered because the
long muon track should be distinguishable from the shower.  Three channels
are considered: CC interaction $\nu_e + n \rightarrow e + X$ and NC interactions 
$\nu_e + n \rightarrow \nu_e + X$ and 
$\nu_\mu + n \rightarrow \nu_\mu + X$, where n stands for nucleon.
Sources of background neutrinos are produced either in the Earth or Sun atmospheres.
The atmospheric flux of neutrinos is taken from \cite{volkova}. An analysis of
the difference between this spectrum and others can be found in \cite{als}. The
solar flux of neutrinos is taken from \cite{ingel}.

The average energy loss in both CC and NC interactions is  
$y = (1-E_{lep}/E_\nu)$ where $E_{lep}$ and $E_\nu$ are the lepton and neutrino energy
respectively. For neutrino energies between 10 and 100 GeV $y$ is about
0.48 gradually 
decreasing to about 0.2 at high energies \cite{gandhi96}. At high energies the
lepton gets about 80\% of the neutrino energy.  In CC interactions, the shower
energy is the sum of all shower products, or $E_\nu$.  In NC interactions, the
shower energy is $E_X = yE_\nu$.

In Figure~\ref{fig:simpcasc} we show the background rates compared to 
the Simpzilla signal. The differential shower rate weighted by shower energy
is shown. We assume no angular resolution and no background reduction. Even in this
worst possible scenario the signal can be detected for a large fraction of the 
Simpzilla parameter space. 

Background reduction will be useful for detecting Simpzilla scenarios 
with low fluxes.  
At very high energy, tau neutrinos can be identified using the double bang
signature \cite{pakvasa} and the zenith angle dependence
\cite{fsaltz}. The double bang signature looks for two showers inside the detector.
The tau will travel a distance $l_\tau$ given by
\begin{equation}
l_\tau = \frac{E_\tau c t_0}{m_\tau} = \frac{(1 - y)E_{\nu_\tau} c t_0}{m_\tau}
\end{equation}
where $E_\tau$ and $m_\tau$ are the tau energy and mass, $E_{\nu_\tau}$ is the
tau neutrino energy, $c$ is the speed velocity and $t_0$ is the tau rest lifetime.
If both showers occur inside the detector this double bang signature can 
distinguish the tau shower from other kind of showers. However for this technique 
to work not only do the showers have to be sufficiently energetic and
separated to be detected individually but also they have to be contained in the detector. 
At 100 TeV, the double bang is only 5 meters separated 
and the two showers will be merged because in ice the scattering length of light 
is about 25 meters, and in water it is about 100 meters.

The second technique \cite{fsaltz} is related to the spectral zenith angle dependence. 
Above an energy of 1 PeV the Earth becomes opaque to electron and muon neutrinos.
The tau regenerates, ie, the tau produced in a neutrino
CC interaction will almost immediately decay back into a tau neutrino. Therefore
the tau neutrino will have its energy degraded when going through the Earth but
at high energies will not have its flux attenuated. The electron or muon
produced in the neutrino CC interaction will be absorbed and the flux will be
attenuated. The tau neutrino flux will have its energy degraded to the
energy with which its interaction length is of the order of the diameter of the 
Earth (which we call transparency energy). In this way the tau neutrino spectrum 
will be a pile up of events around the transparency energy ($\sim 100$ TeV).
Using these two characteristics \cite{fsaltz}
the tau neutrino can be distinguished from the background by the
pile up around 100 TeV. Also it will not have a zenith angle dependence since
the flux will not be degraded. As the electron and muon neutrino have a strong
zenith angle dependence \cite{fsaltz} one can use the flat tau neutrino spectrum
as a signature.

It has also been proposed \cite{bck} that tau neutrinos above $\sim 100$ TeV
will have their signature enhanced by the presence of anti-muon neutrinos and
by anti-electron neutrinos.

The Simpzilla spectrum is softer than 100 TeV, so background reduction in this
analysis must come from the shower properties.  Backgrounds will quickly
dominate below 1 TeV, if energy cuts are not possible.  The IceCube energy threshold
for showers is 1 TeV \cite{pdd}.  This threshold can be raised as needed to
optimize the energy resolution.  Some of the interesting models have many tens
of thousands of events above this threshold.

\begin{figure}
\centering\leavevmode \epsfxsize=450pt \epsfbox{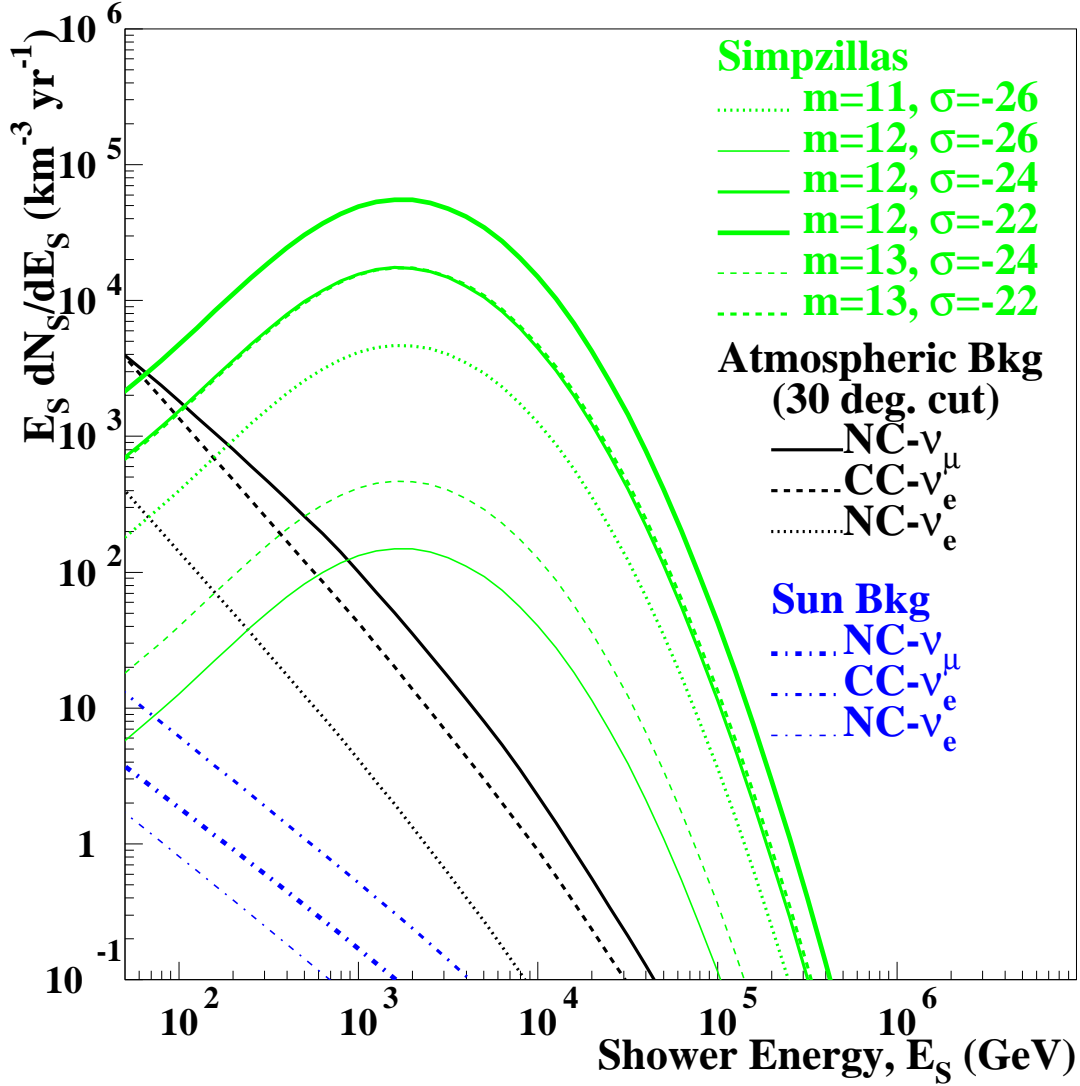}
\caption{Same as Figure~\ref{fig:simpcasc} but now with a $30^o$ angular resolution
used to reduce the atmospheric background.}
\label{fig:backcut}
\end{figure}

Dimmer Simpzilla models will be accessible using angular constraints.  
Energetic showers are boosted along the production direction.  
Charged particles in the shower produce Cherenkov light at broader angles.
This light is also scattered as it travels to the PMTs.  Preliminary studies
indicate that reconstructed shower directions have an RMS angular resolution 
of 30 degrees \cite{pdd}. Figure~\ref{fig:backcut} shows the reduction of the
background by a 30 degree angular cut. One can see that this reduction will
allow a broader Simpzilla parameter space to be probed. 

\subsection{Rate of current and proposed detectors}
\label{sec:simpdet}

As mentioned in the previous section, the atmospheric background for Simpzillas
can be reduced
using a $30^o$ angular cut. The background from the Sun is irreducible but yet very
low when compared to the Simpzilla signal (see Figure~\ref{fig:backcut}).

Although neutrino telescopes
have poor energy resolution it is possible to set a lower energy threshold.
To optimize the energy threshold of detectors with the goal of probing up to a
confidence level (CL) a broader Simpzilla parameter space, we scale the
signal to the background level for a fix set of Simpzilla parameters. This allows 
determination of flux limits and experiments sensitivity as described in Feldman and Cousins
(FC) \cite{feld} and in \cite{pdg}. In order to determine the optimum energy threshold the
systematic uncertainty in the background for each detector has to be understood. This was 
estimated in \cite{als} and results in a 25\% background systematic error for 
a km$3$ detector and 50\% for the smaller experiments as AMANDA-II, ANTARES and NESTOR.
The determination of the experiments sensitivity using
the FC approach is described in \cite{als}. Our results are shown within 95\% CL.
Table~\ref{tab:thres} shows the optimum
energy threshold for the different experiments. The IceCube energy threshold
applies also for an expanded version of ANTARES or NESTOR. The optimum energy threshold
optimizes the background/signal ratio in order to be able to probe a larger Simpzilla
parameter space region.

\begin{table}
\caption[t1]{\label{tab:thres} Optimum energy threshold (in TeV) for Simpzilla detection.
The results considering background from full sky and using a $30^o$ angular cut are shown.}
\begin{tabular}{lcccc}
\hline
 & & 95\% CL & & \\
\hline
 & IceCube (km$3$) & ANTARES & AMANDA-II & NESTOR \\
full sky threshold (TeV) & 20.0 & 6.3 & 4.0 & 3.2 \\
$30^o$ sky (TeV) & 2.5 & 0.5 & 0.3 & 0.25 \\
\hline
 & & $ 5 \sigma$  & & \\
\hline
$30^o$ sky (TeV) & 1.3 & 0.25 & 0.16 & 0.10 \\
\hline
\end{tabular}
\end{table}

The signal and background estimated rate for the smaller experiments is found by 
reducing the
rate by the same factor as for the volume reduction. The AMANDA-II detector has a geometrical 
volume of about $16 \times 10^{-3}$ km$^3$ 
\cite{barwick}, ANTARES \cite{antares} of about $3 \times 10^{-2}$  km$^3$ and NESTOR
\cite{nestor} of about $9 \times 10^{-3}$ km$^{-3}$.
It has been shown \cite{als} that for these telescopes the geometrical efficiency as a function
of zenith angle varies between 100 to 90\% for muons with a path length longer than
100 meters. This will be slightly different for showers but we will assume 100\%
geometrical efficiency for AMANDA-II, ANTARES and NESTOR detectors. 

\begin{figure}
\centering\leavevmode \epsfxsize=450pt \epsfbox{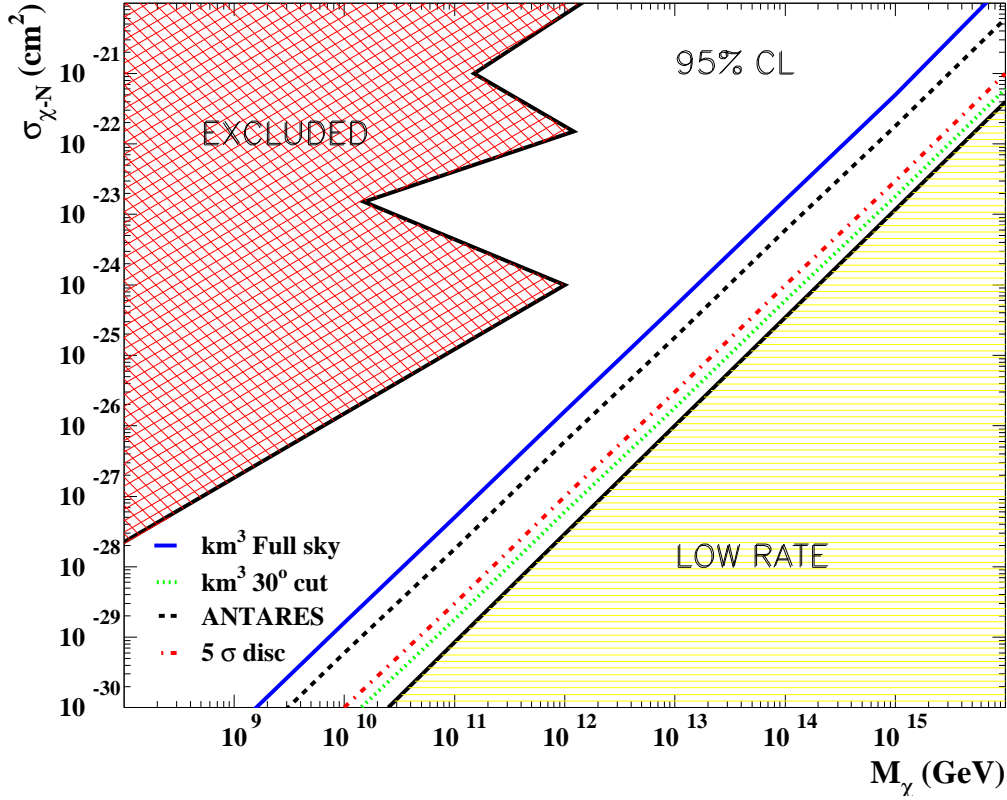}
\caption{The clear region below the jagged line shows the allowed Simpzilla
$M_\chi$ versus $\sigma_{\chi-N}$ space. The hatched region above the jagged line is
excluded by several arguments discussed in \cite{stark}. More model dependent constraints
can be found in \cite{stark,ita}. The lined region indicates
where the neutrino flux is too low \cite{ahk}. The lines show the region to be probed 
with 95\% CL (above the line) in one year of full operation. The solid line
is for a km$^3$ experiment (such as IceCube) with no background reduction and the dotted line 
including
a 30$^o$ angular cut to reduce the background. The dashed line show the region to be probed 
by ANTARES assuming a 30$^o$ angular cut. AMANDA-II and NESTOR limits are between the
ANTARES limit and the full sky IceCube limit. The limits from the smaller telescopes
include a $30^o$ angular cut to reduce the background. The dash-dotted line is the
km$^3$ volume $5 \sigma$ limit. The limits for a km$^3$ volume
detector assumes that the detector is either in the South Pole (IceCube) or
in the Mediterranean sea (an expanded version of ANTARES or NESTOR).}
\label{fig:simpallow}
\end{figure}

Figure~\ref{fig:simpallow} shows the Simpzilla parameter space to be probed by a neutrino
telescope with km$^3$ volume either in the South Pole (IceCube) or in the Mediterranean
(an expanded version of ANTARES or NESTOR). The region excluded (hatched) 
is based on several
experimental measurements \cite{stark} and the ``low rate'' region indicates
where the neutrino flux is too low \cite{ahk}.
The lines show the region to be probed (above the line) within 95\% CL. The solid line
is for a km$^3$ experiment with no background reduction and the dotted line including
a 30$^o$ angular cut to reduce the background. The smaller experiments must use
a technique to reduce the background otherwise they cannot probe a significant
Simpzilla parameter region. The dashed line shows the region to be probed by ANTARES 
assuming a 30$^o$ angular cut. Among the smaller experiments it can
probe the largest parameter space. AMANDA and NESTOR will be similar and their limits are 
between the ANTARES and full sky km$^3$ detector limits. This figure also 
shows the parameter space
(dashed-dotted) for a $5 \sigma$ detection which assures discovery if Simpzillas
are the dark matter. This
limit is for a km$^3$ volume assuming a 30$^o$ angular cut. ANTARES discovery limit
is close to the km$^3$ full sky limit.

In the most natural scenarios the Simpzilla mass is close to the inflaton
mass ($10^{12}$ GeV) and Simpzilla--Nucleon cross section is close to the Nucleon--Nucleon
strong interaction cross-section. As can be seen from Figures~\ref{fig:backcut} 
and \ref{fig:simpallow} these can be easily probed by neutrino telescopes of
km$^3$ volume. Also the smaller neutrino telescopes can probe a large region of the
Simpzilla parameter space.

\subsection{Neutrino Oscillations}
\label{sec:osc}

There is strong evidence that neutrinos oscillate from one flavor to another.
Superkamiokande \cite{sk} determines the most probable solution for 
the difference between the squared masses of the two neutrino mass
eigenstates ($\Delta m^2$) as $3.5 \times 10^{-3}$ eV$^2$ and a large mixing angle
for atmospheric neutrinos. It also favors muon neutrino oscillation into tau 
neutrinos \cite{sktau}. SNO \cite{sno} has recently determined $\Delta m^2$ to be 
$5.0 \times 10^{-5}$ eV$^2$ in the large mixing angle scenario for solar neutrinos.
Atmospheric neutrino oscillations will be important only below $\sim 200$ GeV
\cite{as}. This is below the energy threshold for the Simpzilla analysis and therefore
has no effect in the background considered here. However neutrino oscillations
might occur in the Sun and in transit to Earth and therefore affect the Simpzilla 
signature. 
One should note that the neutrinos from Simpzilla annihilation have energies much 
higher than the standard solar neutrinos. Oscillation of three neutrino flavors are
analyzed in the GeV energy regime \cite{andre}. One important result is that
for some regions of the oscillation parameter space (which now includes three mixing
angles) it is possible to have an enhancement of tau neutrinos detected in the
Earth (when coming from the Sun) with respect to the number of muon neutrinos 
(or vice-versa).

Crotty \cite{crotty} analyses the effect of oscillations in the neutrino flux
from Simpzilla annihilation in the Sun assuming fixed values for the three flavor
oscillation parameters. $\Delta m^2_{31}$ is taken as $3 \times 10^{-3}$ eV$^2$,
$\sin^2 \xi = 0.1$, $\sin^2 \theta = 0.5$, $\Delta m^2_{21} = 2 \times 10^{-5}$ eV$^2$
and $\sin^2 \omega = 0.2$. 
It is also assumed the ``normal'' mass hierachy where $ m_1 < m_2 < m_3$.
See \cite{crotty} for a description of these parameters.
The result is that the tau neutrino estimated rate at the Earth and the
tau neutrino contained rate in a km$^3$ volume will be reduced
while the muon neutrino event rate will be enhanced. It has also been shown \cite{bck} 
that secondary muon neutrinos (see section~\ref{sec:simp}) will enhance the
muon neutrino flux by about 20\% of the tau neutrino flux at the Earth.
Taking both secondaries and oscillation into account, the tau neutrino and muon
neutrino flux at the Earth become approximately equal \cite{crotty}. The
conversion rate in a km$^3$ volume will also be approximately equal (with a
slightly higher muon neutrino rate) \cite{crotty}. The tau neutrino detected
rate will then be approximately half of that without oscillations.
 
This reduction in the tau neutrino rate does not reduce the Simpzilla
parameter space to be probed by neutrino telescopes which are also able to
detect muon neutrinos. The region to be probed by
tau neutrino detection would be reduced. The km$^3$ with $30^o$ sky limit in 
Figure~\ref{fig:simpallow} would be moved to where the 5 sigma limit stands without
oscillation. The ANTARES limit will move very close to the km$^3$ full sky limit.
The limit that will be most affected will be the $5 \sigma$ which   
will also move close to the km$^3$ full sky limit. However once oscillations are
considered there will be a muon neutrino flux at the same level of the tau
neutrino flux. As mentioned before (see section \ref{sec:simprate}) the neutrino
telescopes angular resolution is much better for muons (about 1 degree) than for
the tau generated showers. This will reduce the background to a much lower level
than for the tau neutrino analysis. As a result the Simpzilla parameter space to
be probed is the same as without oscillations or even larger.
This is subject of future work.

\section{Conclusion}

We have discussed the indirect detectability of dark matter by neutrino telescopes. 
The decay of secondary products of dark matter annihilation would generate a flux of
high energy neutrinos.

We have argued that when current and proposed neutrino telescopes' energy threshold and 
location are taken into
account the detection of the neutralino as a WIMP is borderline.
Most of the WIMP parameter
space is inaccessible even in tens of years of observation.
The parameter space which can be probed in 5 or more years of 100\% live-time
can also be probed by direct detection. These two different techniques will complement
each other.

However if the dark matter is constituted by superheavy (mass 
$\mathrel{\vcenter{\hbox{$>$}\nointerlineskip\hbox{$\sim$}}} 10^{10}$ GeV) dark matter
(Simpzillas), most of the current and proposed neutrino telescopes can probe the most natural
scenarios. In the most natural scenario, the Simpzilla mass is similar to the inflaton
mass ($10^{12}$ GeV) and the Simpzilla--Nucleon cross section is the same as the
strong interaction cross section ($10^{-26}$ cm$^2$).

Proposed detectors such as IceCube or an expanded version of ANTARES or
NESTOR are able to probe Simpzillas to $5 \sigma$ having a significant
signal rate. 
Figure~\ref{fig:simpallow} shows the Simpzilla mass versus Simpzilla - Nucleon cross-section
region to be probed. A 30$^o$ angular cut efficiently
reduces the background without any loss in the signal rate. This cut allows
most of the currently allowed Simpzilla parameter space to be probed.

If oscillations are taken into account tau neutrinos might convert into muon
neutrinos. Crotty \cite{crotty} shows that under certain parameter assumptions
this might reduce the tau neutrino detected rate by about a factor of two and
enhance the muon neutrino detected rate. As these muon neutrinos can be detected,
not only the Simpzilla parameter
space to be probed will continue to be as shown in Figure~\ref{fig:simpallow}
but it might even allow for neutrino oscillation studies.

The smaller neutrino telescopes, AMANDA-II and current versions of ANTARES and
NESTOR, are able to probe a large fraction of Simpzilla mass versus Simpzilla--Nucleon 
cross-section parameter space. 

As a final conclusion, although WIMP indirect detection 
of most models will take
tens of years, current and proposed neutrino telescopes can either discover or rule
out the most natural scenarios for superheavy strongly interacting massive particles 
in one year of full operation.

\begin{acknowledgements}
We thank Joakim Edsjo and Willi Chinowsky for comments.

This work was supported by NSF Grants KDI 9872979 and Physics/Polar Programs 
0071886
and in part by the Director, Office of
Energy Research, Office of High Energy and Nuclear Physics, Division of
High Energy Physics of the U.S. Department of Energy under Contract No.
DE-AC03-76SF00098 through the Lawrence Berkeley National Laboratory.
\end{acknowledgements}

\end{document}